\documentclass[conference]{IEEEtran}
\IEEEoverridecommandlockouts
\newtheorem{theorem}{Theorem}
\newtheorem{corollary}[theorem]{Corollary}
\newtheorem{lemma}[theorem]{Lemma}

\newtheorem{definition}[theorem]{Definition}

\usepackage[noadjust]{cite}

\ifCLASSINFOpdf
  \usepackage[pdftex]{graphicx}
\else
\fi

\usepackage[cmex10]{amsmath}
\usepackage{amssymb}

\usepackage{enumitem}

\usepackage[caption=false,font=footnotesize]{subfig}

\begin{document}
\title{On the Girth of (3,L) Quasi-Cyclic LDPC Codes based on Complete Protographs\thanks{This material is based upon work supported by the Broadcom Foundation and the National Science Foundation under Grant Numbers 1162501 and 1161822. Any opinions, findings and conclusions or recommendations expressed in this material are those of the author(s) and do not necessarily reflect the views of the National Science Foundation.  This research was carried out in part at the Jet Propulsion Laboratory,
California Institute of Technology, under a contract with NASA and JPL-NSF Task Plan 82-17473.  }}

\author{\IEEEauthorblockN{Sudarsan V. S. Ranganathan\IEEEauthorrefmark{1}, Dariush Divsalar\IEEEauthorrefmark{2} and Richard D. Wesel\IEEEauthorrefmark{1}}
\IEEEauthorblockA{\IEEEauthorrefmark{1}Department of Electrical Engineering, 
University of California, Los Angeles, Los Angeles, California 90095\\ \IEEEauthorrefmark{2}Jet Propulsion Laboratory, California Institute of Technology, Pasadena, California 91109\\
Email: sudarsanvsr@ucla.edu, Dariush.Divsalar@jpl.nasa.gov, wesel@ee.ucla.edu}
}


\maketitle
\begin{abstract}
We consider the problem of constructing $(3,L)$ quasi-cyclic low-density parity-check (LDPC) codes from complete protographs. A complete protograph is a small bipartite graph with two disjoint vertex sets such that every vertex in the variable-node set is connected to every vertex in the check-node set by a unique edge.  This paper analyzes the required lifting factor for achieving girths of six or eight in the resulting quasi-cyclic codes with constraints on lifting.  The required lifting factors provide lower bounds on the block-length of such codes. 
\end{abstract}
\IEEEpeerreviewmaketitle
\section{Introduction and Background}
\label{sec_introduction_full}
{\em Protograph-based quasi-cyclic LDPC codes} (protograph QC-LDPC)~\cite{Proto,1317123} are LDPC codes~\cite{ldpcgall} with encoders and decoders amenable to implementation for practical purposes. Generally, a code constructed from a protograph need not be quasi-cyclic. A QC code is built from a protograph by restricting the permutation matrices used in the lifting process to be circulants.  A protograph QC-LDPC code can be described by specifying the permutation shift indices of the circulant permutation matrices associated with the lifting process~\cite{1317123}.  

A protograph~\cite{Proto} defines the family of codes that can be obtained from it by lifting and many properties of the codes in the family depend on the graphical structure of the chosen protograph. In this paper, we consider the case where the protograph is a {\em simple} (has no loops or multiple links between two vertices) and {\em complete} (every vertex in the variable-node set is connected to every vertex in the check-node set) bipartite graph. QC-LDPC codes obtained from simple and complete protographs are called {\em conventional} QC-LDPC codes in~\cite{6475181}, which considers simple QC-LDPC codes in general, including the subset which are conventional.

The performance of LDPC codes is dictated, to a certain extent, by the girth of the codes. Also, in the regime of short-to-moderate block-lengths, the minimum distance of an LDPC code affects its performance in the error-floor region if the variable-node degrees are small~\cite{Paper1}. In this regard, the minimum distance of a protograph QC-LDPC code and its girth are interrelated as suggested by the work in~\cite{6145509}. The works in~\cite{6475181,1317123} derive lower bounds on the required lifting factors (and thus block-lengths) for obtaining various girths for QC codes constructed from protographs and provide the foundations for this paper.  Works including~\cite{1317123,6475181,6587468,4658694,isita2006_qcmin} have focused on obtaining these bounds because they are of practical importance and have demonstrated code construction techniques to obtain codes with as high a girth as possible. 

We focus on the case of $(3,L)$ protograph QC-LDPC codes. These are regular codes that perform well over many rates. The paper is organized as follows: Section~\ref{sec_notation} introduces notation. Section~\ref{sec_min_6} completely characterizes the lifting requirements to construct a $(3,L)$ code with girth 6 when the lifting factor is equal to $L$ and gives an explicit construction that achieves a girth of 6 for any possible value of $L$. Section~\ref{sec_girth8} derives a bound (under a constrained setting) on the lifting factor required to obtain a girth of at least 8.  This bound improves on the bounds in~\cite{6475181,1317123}. Section~\ref{sec_conclusion} concludes the paper. 
\section{Definitions and Notation}
\label{sec_notation}
A {\em protograph}~\cite{Proto} is a small bipartite Tanner graph~\cite{Tanner} and a {\em protomatrix} is a biadjacency matrix of the protograph. A graph's {\em girth} is the length of its shortest cycle(s). The protomatrices considered in this paper have the form
\begin{align}
H_{\text{protomatrix}}=\begin{bmatrix}
1 & 1 & 1 & \cdots \\
1 & 1 & 1 & \cdots \\
1 & 1 & 1 & \cdots \\
\end{bmatrix}_{3 \times L}.
\end{align}
At places, the terms protograph and protomatrix are used interchangeably. Associated with any protomatrix, the process of {\em lifting} to obtain a QC code is the replacement of every non-zero entry $z$ in the protomatrix by a sum of $z$ circulant permutation matrices (CPMs) of size $N \times N$ with distinct support and every 0 in the protomatrix by an $N \times N$ matrix of all zeros. If the protomatrix is of size $J \times L$ then lifting yields a parity-check matrix $H$ of size $JN \times LN$.  Because our protomatrices are simple and complete, lifting replaces every entry in the protomatrix with an $N \times N$ CPM. 

\begin{definition}[Permutation-shift matrix~\cite{6475181}]
\label{def_psm}
The permutation-shift matrix $P$ of a QC-LDPC code constructed from a $J \times L$ protomatrix with entries at most equal to 1 is the $J \times L$ matrix of permutation shift indices that are chosen for the non-zero entries of the protomatrix during the process of lifting. With the lifting factor being $N$, an element $0 \le x \le N-1$ in $P$ corresponds to a CPM in the parity-check matrix $H$ obtained via $x$ circular shifts of the rows of the identity matrix of size $N \times N$.  The orientation (left or right) of the permutation shifts is unspecified in this paper without loss of generality (WLOG). 
\end{definition}

The cyclic group of integers modulo $N$, $\{0, 1, \dots, N-1\}$, is denoted $\mathbb{Z}/N$. This is the set of first $N$ non-negative integers with addition modulo-$N$ as the associated binary operation, represented by $x_i + x_j$.  Similarly,  $x_i - x_j = x_i + \left(-x_j\right)$ represents adding the inverse of $x_j$ to $x_i$. The {\em order} of a group is its cardinality. A {\em permutation} $\pi$ is a bijective map of a finite set of elements onto itself. 

\begin{definition}
\label{def_fp}
A permutation $\pi$ of $\mathbb{Z}/N$ is said to have a {\em fixed point} if $\pi(i) = i$ for any $i=0, 1, \dots, N-1$.
\end{definition}
\section{On the Minimum Lifting Factor for Girth Greater Than or Equal To 6}
\label{sec_min_6}
We consider the special case of this problem with the constraint that the lifting factor $N$ satisfies $N=L$. This is the least value of $N$ for which one can possibly obtain a girth of $g>4$~\cite{1317123,6475181}. By looking at this special case we arrive at a combinatorial interpretation to the problem of obtaining codes with girth at least 6 from complete protomatrices of size $J \times L$. Since $N=L$, we may use $N$ and $L$ interchangeably. 

Works including~\cite{1317123} have constructed codes via computer searches to show empirically the existence of codes with girth $g \ge 6$ for some odd values of $N=L$ (including analytical constructions for all primes; see~\cite{fan_array} also). We show analytically that for all odd values of $N=L$, there exist $(3,L)$ codes with girth $g \ge 6$. \cite{isita2006_qcmin} has established this result and our contribution is a proof via combinatorial structures called \emph{complete mappings}~\cite{paige1947}. We provide an algebraic construction that produces codes with girth $g = 6$ for any odd $N=L$. This construction includes, as a special case, the array-code based proof of \cite{isita2006_qcmin} for the $(3,L)$ case. 

\begin{lemma}[\cite{1317123}]
\label{lem_main_lemma}
With the lifting factor being $N$, in any QC-LDPC code with a protomatrix with no entry larger than 1, a cycle of length $\ell$ ($\ell$ even) in the Tanner graph of the code can be equivalently described by a sequence of edges $(e_1, e_2, \dots, e_{\ell})$ in the protograph whose corresponding permutation shifts in $P$ that are given as $x_1, x_2, \dots, x_\ell$ satisfy
\begin{align}
\label{e6}
\sum_{i=1}^{\ell}(-1)^{i+1}  x_i = 0 \mod N,
\end{align}
where $e_i \ne e_{i+1}$ for all $i \in \{1, 2, \dots, \ell-1\}$ and $e_1 \ne e_\ell$. Consecutive pairs of consecutive edges $\{e_i, e_{i+1}\}$ for all $i \in \{1, 2, \dots, \ell-1\}$ and $\{e_{\ell}, e_1\}$ alternatingly lie in the same row or same column of the protomatrix. 
\end{lemma}

The elements of $P$ are assumed to be in $\mathbb{Z}/N$ and thus ``mod $N$" may not be mentioned at most places that involve operations with elements from $P$.

\begin{lemma}[Extension of~\cite{1317123}, Theorem 2.2]
\label{lem_extension}
With a lifting factor of $N=L$, any permutation-shift matrix $P$ that could lead to $g>4$ for a $(3,L)$ code with a complete protomatrix may be written WLOG as
\begin{align}
\label{e1}
P = \begin{bmatrix}
0 & 0 & 0 & \cdots & 0 \\ 0 & 1 & 2 & \cdots & N-1 \\ \pi(0)=0 & \pi(1) & \pi(2) & \cdots & \pi(N-1)
\end{bmatrix},
\end{align}
where $\pi$ has only one fixed point at $\pi(0) = 0$.
\end{lemma}

\begin{IEEEproof}
Irrespective of the $3L$ indices that are chosen for $P$, one can always apply circular shifts to the row blocks and the column blocks of $H$ (after lifting) to obtain an isomorphic graph for which the first row and column have all-zero indices in $P$, as observed in~\cite{1317123}. For girth $g>4$, \cite{1317123} shows that no non-zero element can repeat in the same row or the same column. Thus the non-zero entries in each of rows 2 and 3 are all unique within the respective rows and the ordering of row 2 in~\eqref{e1} can be obtained WLOG by rearranging the columns once we have 0's in row 1 and column 1. To ensure that no column repeats a nonzero value, the permutation cannot have any fixed point except $\pi(0) =0$.
\end{IEEEproof}

The preceding lemma implies that, WLOG, only $L-1$ non-zero permutation shifts need to be specified and these belong to the third row. As an example where repetition in the same column prevents $g>4$, the case of $L=2$ leads to $g=4$ as there is only one non-zero element in $\mathbb{Z}/2$. The probability that a permutation of a finite number of elements ($N-1$) has no fixed points asymptotically, as $N \rightarrow \infty$, equals $\frac{1}{e}$~\cite{matousek2008invitation}.  If we search randomly for permutations of $N-1$ non-zero elements to achieve a girth of $g>4$, then the number of permutations to be considered when constructing a code for large values of $N-1$ is very high but only approximately 36.8\% of them will pass the preliminary test of not having a fixed point. 

\begin{definition}[Complete mapping~\cite{paige1947,oeisA003111}]
\label{def_complete_mapping}
A complete mapping of the cyclic group $(\mathbb{Z}/N,+)$ is a permutation $\pi$ which satisfies $\pi(0)=0$ and that $(0, \pi(1) - 1, \pi(2)-2, \dots, \pi(N-1)-(N-1))$ is also a valid permutation. 
\end{definition}

\begin{theorem}
\label{thm_main}
With a lifting factor of $N=L$, the parity-check matrix $H$ of a code with a complete protomatrix of size $3 \times L$ has a girth $g>4$ if and only if the permutation $\pi$ of $\mathbb{Z}/N$ that specifies the third row of $P$ in~(\ref{e1}) is a complete mapping. 
\end{theorem}

\begin{IEEEproof} 
Consider any two columns of the shift matrix of~(\ref{e1}) and form a $2 \times 2$ sub-matrix of rows 2 and 3 out of the chosen columns as
\begin{align*}
\begin{bmatrix}
x_i & x_j \\ x_k & x_{\ell}
\end{bmatrix},x_k = \pi(x_i), x_{\ell} = \pi(x_j).
\end{align*}
From the general condition of~\eqref{e6} in Lemma~\ref{lem_main_lemma},  $x_i, x_j, x_k, x_{\ell}$ lead to cycle(s) of length four if and only if (iff)
\begin{align}
\label{e2}
x_i-x_k+x_{\ell}-x_j = 0.
\end{align}
Rewriting the above, the girth is greater than 4 iff
\begin{align}
\label{e4}
\left(x_{\ell}-x_j\right)-(x_k-x_i) \ne &  0,
\end{align}
which means that $x_{\ell}-x_j \ne x_k-x_i$ should be satisfied for any $x_i,x_j,x_k,x_{\ell}$ as considered above. This is possible iff 
\begin{align}
(\pi(\text{row 2}) - \text{row 2})
\end{align}
describes a permutation (i.e.\ the sequence contains each distinct element in the group exactly once), which occurs iff row 3 is a complete mapping.
\end{IEEEproof}

\begin{theorem}
\label{thm_res1}
There exists a $(3,L)$ quasi-cyclic LDPC code with a complete protograph lifted by a factor $N=L$ satisfying girth $g>4$ iff $L$ is odd. 
\end{theorem}

\begin{IEEEproof}
From~\cite{paige1947}, there exists a complete mapping of a finite abelian group of order $N$ iff the group does not possess exactly one element of order 2. When $N$ is even, this condition is violated as one can verify that $\frac{N}{2}$ is the only order-2 element in the finite abelian group $\mathbb{Z}/N$. On the contrary, in finite groups $\mathbb{Z}/N$ of odd orders there exists no element of order 2, according to Lagrange's theorem on the order of elements in a finite group. This argument in conjunction with Theorem~\ref{thm_main} completes this proof.
\end{IEEEproof}

The number of complete mappings of $\mathbb{Z}/N$ is documented in~\cite{oeisA003111}. The first few terms of this sequence as a function of $N$, from $N=1, 3, 5, \dots$, are $1$, $1$, $3$, $19$, $225$, $3441$, $79259$, $2424195$, $94471089$, $4613520889$.  For odd $N=L$ all the complete mappings that yield codes with girth $g \ge 6$ lead to $g=6$ since girth $g \ge 8$ requires a higher lifting factor (see Section~\ref{sec_girth8}). For odd $N=L$, random search might identify a complete mapping and hence a $g=6$ code, but the probability of any randomly selected mapping being complete decreases quickly with increasing $L$. For instance, when $L=15$ corresponding to a design rate $R=\frac{L-3}{L}=0.8$ this probability is $\frac{2424195}{14!} = 0.000028$ and when $L=17$ and $R=0.8235$ this probability is $0.000004$ and so on. In the following we present a family of complete mappings and thus a family of codes for any odd $N=L, L \ge 3$ that have $g = 6$.

\begin{corollary}[Product construction]
\label{cor_res1_1}
Consider the following mapping for row 3 in~(\ref{e1}) with $h \in \{2, 3, \dots, N-1\}$:
\begin{align}
\label{e5}
\pi_p(i) = hi \mod N, 0 \le i \le N-1,
\end{align}
where $hi \mod N$ is multiplication modulo-$N$ of integers $h$ and $i$. For $N=L$ odd and $N \ge 3$, if $h$ and $h-1$ are each coprime with $N$, then $\pi_p$ is a complete mapping of $\mathbb{Z}/N$ and thus leads to a $(3,L)$ code with girth 6.
\end{corollary}

\begin{IEEEproof}
Note that since $h$ is chosen to be coprime with respect to $N$, $(hi \mod N:0 \le i \le N-1)$ is a valid permutation of $\mathbb{Z}/N$. This is because $hi-hj = h(i-j) \ne 0 \mod N, \forall i \ne j$ as $h$ is not a factor of $N$. We need to further show that~(\ref{e2}) from Theorem~\ref{thm_main} has no solution. Writing~(\ref{e4}), which is obtained from~(\ref{e2}), for this permutation:
\begin{align}
&(x_{\ell}-x_j)-(x_k-x_i) \ne  0 \notag \\
\iff &h(x_j-x_i)-(x_j-x_i) \ne  0, \text{ as $x_k = hx_i, x_\ell = hx_j$ } \notag \\
\iff &(h-1)(x_j-x_i) \ne  0, \notag
\end{align}
which is satisfied for this permutation for all $x_j\ne x_i$ since $h-1 \ge 1$ is chosen to be coprime with respect to $N$. 
\end{IEEEproof}

There exists such an $h$ for every odd $N \ge 3$.  An example is $h=N-1$ for which
\begin{align}
\label{e3}
P = \begin{bmatrix}
0 & 0 & 0 & \cdots & 0 & 0\\ 0 & 1 & 2 & \cdots & N-2 & N-1 \\ 0 & N-1 & N-2 & \cdots & 2 & 1
\end{bmatrix} \, .
\end{align}

Also, $(3,L)$ array codes~\cite{fan_array,1317123}, for any odd $L \ge 3$ (not necessarily prime), are a special case of the preceding construction with $h=2$ and thus have $g=6$ \cite{isita2006_qcmin}.

\begin{corollary}
\label{cor_res1_2}
If $N=L$ is even then there exists a $(3,L)$ complete-protomatrix-based code with girth equal to 4 whose Tanner graph has exactly $N$ cycles of length four.
\end{corollary}

\begin{IEEEproof}
This follows from~\cite{paige1947}, which proves that in case the order of a finite abelian group is even then there exists an ``almost complete" mapping $\pi$ of the group such that the sequence $(0, \pi(1) - 1, \pi(2)-2, \dots, \pi(N-1)-(N-1))$ has exactly $N-1$ distinct elements. Thus, one element appears twice. This implies that there exists a mapping for the third row such that only one $2 \times 2$ block from the second and the third rows leads to $N$ length-4 cycles.
\end{IEEEproof}

As shown in \cite{isita2006_qcmin}, it can also be observed that if $L$ is even then there exists a $(3,L)$ complete-protomatrix-based code with a girth of 6 if the lifting factor is $N = L+1$. 


One can generalize the discussion so far to see that for the $(J,L)$ case there could be a code with $g>4$ when the lifting factor is $N=L$ only if there exist $J-2$ distinct complete mappings of $\mathbb{Z}/N$. This condition is necessary but not sufficient because the rows produced by the $J-2$ complete mappings also have to satisfy the following condition:   Every pair of the ${J-2 \choose 2}$ rows indexed by $\{\{i,j\}:3 \le i < j \le J\}$ are such that row $j$ is a complete mapping of row $i$. 

Consider the computer-search based Table~\ref{table_fossorier} of~\cite{1317123} (reproduced below). When $N=L=9$, the computer search could not find a $(J,9)$ code with girth $g=6$ when $J \ge 4$.  Using the previous paragraph, we can confirm that such a code does not exist.  There are $225$ complete mappings of $\mathbb{Z}/9$. We can corroborate the  result in this table since not even one pair out of ${225 \choose 2}$ pairs of complete mappings can satisfy the requirement that one row in the pair is a complete mapping of the other. 

\begin{table}[h]
\renewcommand{\arraystretch}{1.1}
\caption{Smallest Value of $N$ for which a $(J,L)$ Code with Girth $g \ge 6$ was Found in~\cite{1317123} Using Computer Search}
\label{table_fossorier}
\centering
\begin{tabular}{||c||c|c|c|c|c|c|c|c|c||}
\hline
  $~~~~L$&$4$&$5$&$6$&$7$&$8$&$9$&$10$&$11$&$12$\\
  $J~~~~$&&&&&&&&&\\
\hline
\hline
  $3~~~~$&$5$&$5$&$7$&$7$&$9$&$9$&$11$&$11$&$13$\\
\hline
  $4~~~~$&$-$&$5$&$7$&$7$&$9$&$\mathbf{10}$&$11$&$11$&$13$\\
\hline
  $5~~~~$&$-$&$-$&$7$&$7$&$9$&$\mathbf{10}$&$11$&$11$&$13$\\
\hline
\end{tabular}
\end{table}
\section{Towards a Tighter Bound on the Required Lifting Factor for Girth $\ge$ 8 while $L \ge 4$}
\label{sec_girth8}
Assuming $L \ge 4$, it is known that the lifting factor $N$ has to satisfy $N > 2(L-1)$ to obtain a girth of $g \ge 8$ for our $(3,L)$ codes~\cite{1317123}. In this section, we derive an improved bound on this required lifting factor under a constraint by using an additive combinatorics formulation of the problem. It is conjectured, for future investigation, that the bound holds without this imposed constraint. 

The following lemma states the necessary and sufficient conditions  of~\cite{1317123} for the permutation-shift matrix $P$ of a complete-protomatrix-based $(3,L)$ code to achieve $g\ge8$. 

\begin{lemma}
\label{lem_main_girth8}
For $L \ge 4$, let $L' = L-1$ and the lifting factor be $N$. The permutation-shift matrix
\begin{align}
\label{e10}
P = \begin{bmatrix} 0 & 0 & 0 & \dots & 0 \\ 0 & x_{1} & x_{2} & \dots & x_{L'} \\ 0 & x_{L'+1} & x_{L'+2} & \dots & x_{2L'}
\end{bmatrix}
\end{align}
leads to a girth of $g \ge 8$ iff all the following conditions hold: With $i, j \in \{1, 2, \dots, 2L'\}$,
\begin{enumerate}
\item $x_i \ne x_j$ for all $i \ne j$ and $x_i \ne 0$ for all $i$.
\end{enumerate}
Fixing $i \ge L' + 1$ and $j = i-L'$ (so that $x_i$ and $x_j$ are in the same column of $P$, with $x_i$ in the third row):
\begin{enumerate}[resume]
\item $x_i - x_j \ne -x_k$, where $k \in \{1, 2, \dots, L'\} \setminus \{j\}$,
\item $x_i - x_j \ne x_k$, where $k \in \{L'+1, L'+2, \dots, 2L'\} \setminus \{i\}$,
\item $x_i - x_j \ne x_k - x_{\ell}$, where $k \in \{L'+1, L'+2, \dots, 2L'\} \setminus \{i\}, \ell \in \{1, 2, \dots, L'\} \setminus \{j\}, k \ne \ell + L'$,
\item $x_i - x_j \ne x_k - x_{\ell}$, where $k \in \{L'+1, L'+2, \dots, 2L'\} \setminus \{i\}, k = \ell + L'$.
\end{enumerate}
\end{lemma}

\begin{IEEEproof}
Condition 1 is Theorem $2.4$ of~\cite{1317123}, which yields the necessary condition $N>2(L-1)=2L'$ for achieving $g \ge 8$. Conditions 2 and 3 apply~(\ref{e6}) to the first column and any other two columns of the shift matrix in~\eqref{e10}. Condition 4 similarly considers any three columns apart from the first (all-zeros) column. Condition 5 avoids length-4 cycles from rows 2 and 3 of $P$. 
\end{IEEEproof}

\begin{definition}[Girth-8 table]
\label{def_g8t}
A {\em girth-8 table} ($G_8$ table) of a $(3,L)$ complete-protomatrix-based QC-LDPC code whose permutation-shift matrix is $P$, using the notation of Lemma~\ref{lem_main_girth8}, is a table of $L'\times L'$ differences:

\begin{table}[h]
\begin{center}\small
\begin{tabular}{c || c | c | c | c} 
+\textbackslash - & $x_{1}$ & $x_{2}$ & \dots & $x_{L'}$ \\ \hline \hline
$x_{L'+1}$ & $d_1 = x_{L'+1} - x_1$ & $x_{L'+1} - x_2$ & \dots & $x_{L'+1} - x_{L'}$ \\
$x_{L'+2}$ & $x_{L'+2} - x_1$ & $d_2$ & \dots & $x_{L'+2} - x_{L'}$ \\
\vdots & \vdots & \vdots & $\ddots$ & \vdots \\
$x_{2L'}$ & $x_{2L'} - x_1$ & $x_{2L'} - x_2$ & \dots & $d_{L'}$
\end{tabular}
\end{center}
\end{table}
A \emph{valid} $G_8$ table is one which leads to a girth of $g \ge 8$.
\end{definition}

\begin{lemma}
A $G_8$ table is valid iff
\begin{enumerate}
\item The set of row and column headers together has $2L'$ distinct non-zero elements,
\item The diagonal elements $d_1, d_2, \dots, d_{L'}$ are all different from the inverses of the column headers, 
\item The diagonal elements are all different from the row headers,
\item None of the diagonal elements is equal to any of the off-diagonal elements of the table,
\item The diagonal elements are all distinct. 
\end{enumerate}
\end{lemma}

\begin{IEEEproof}
These conditions are the equivalent conditions of Lemma~\ref{lem_main_girth8} in the same order. Note that a valid $G_8$ table has no 0 anywhere in it. Conditions 4 and 5, which are mathematically the same albeit for the choice of elements involved but stated separately for clarity, according to Lemma~\ref{lem_main_girth8}, justify uniquely identifying the diagonal elements as $d_1, d_2, \dots, d_{L'}.$
\end{IEEEproof}

\begin{theorem}
\label{thm_main_girth8}
Let the $L'$ rows of any valid $G_8$ table be considered as sets of $L'$ elements each and denoted $A_1, A_2, \dots, A_{L'}$. If there exist two rows $i\ne j$ such that $|A_i \cap A_j| = 0$ or $|A_i \cap A_j| = L'-1$ then such a valid $G_8$ table corresponds to a lifting factor of $N \ge 3L' - 1$. 
\end{theorem}

\begin{IEEEproof}
In general, $|A_i \cap A_j| \le L'-1, i \ne j$ since every row has a diagonal element that is distinct from the elements in the rest of the table. The proof, which is given in the rest of this section, applies conditions 1, 4 and 5 from Lemma~\ref{lem_main_girth8}.

The case where $\exists ~i \ne j: |A_i \cap A_j|=0$ is considered first. If so, then $|A_i|+|A_j|=2L'$ and the rest of the $L'-2$ rows contribute at least one distinct element each as their diagonal elements have to be distinct and thus the number of distinct non-zero elements is at least $3L'-2$ and $N \ge 3L'-1$.

For the second case, assume WLOG that the rows $i,j$ are the first two rows of the $G_8$ table, corresponding to $A_1$ and $A_2$, or the table can be rearranged accordingly (this corresponds to permuting the columns of $P$). Denote the $L'$ distinct elements of $A_1$ (in order from left to right) as
\begin{align*}
d_1 = x_{L'+1} - x_1, f_{1}, f_{2}, \dots, f_{{L'-1}}.
\end{align*}
Any $A_i, i \ne 1$ can be derived from $A_1$ through an offset.  For example, $A_2$ can be obtained from $A_1$ by adding $\Delta = x_{L'+2} - x_{L'+1}$ to $d_1, f_1, f_2, \dots, f_{L'-1}$ in that order. 

The supposition $|A_1 \cap A_2| = L'-1$ implies that $A_1, A_2$ differ in only $d_1 \ne d_2$. Since $d_1$ does not repeat or ``lead to" a new element (or else $|A_1 \cap A_2| < L'-1$), while adding $\Delta \ne 0$ to it and since the only new element that is formed in this second row is $d_2 = f_1 + \Delta$, this means that $d_1 + \Delta = f_i$ for some $i \in \{1,2, \dots, L'-1\}$. Also $\forall k \in \{2, 3, \dots, L'-1\}$ there exists a unique $\ell_k \in \{1, 2, \dots, L'-1\} \setminus \{k\}$ such that $f_k + \Delta = f_{\ell_k}, f_k + \Delta \ne d_1, f_k + \Delta \ne d_2$. 

\begin{definition}[Circular representation]
\label{def_circ_rep}
We choose to represent the elements of $\mathbb{Z}/N$ as unique points on a circle in order from 0 through $N-1$ in the anticlockwise direction such that $N-1$ appears on the circle before crossing 0 when counting from 0 (as integers). With this representation, addition corresponds to moving along the anticlockwise direction. 
\end{definition}

\begin{figure}[h]
\centerline{\subfloat[Without wrap around]{\includegraphics[width=1.43in]{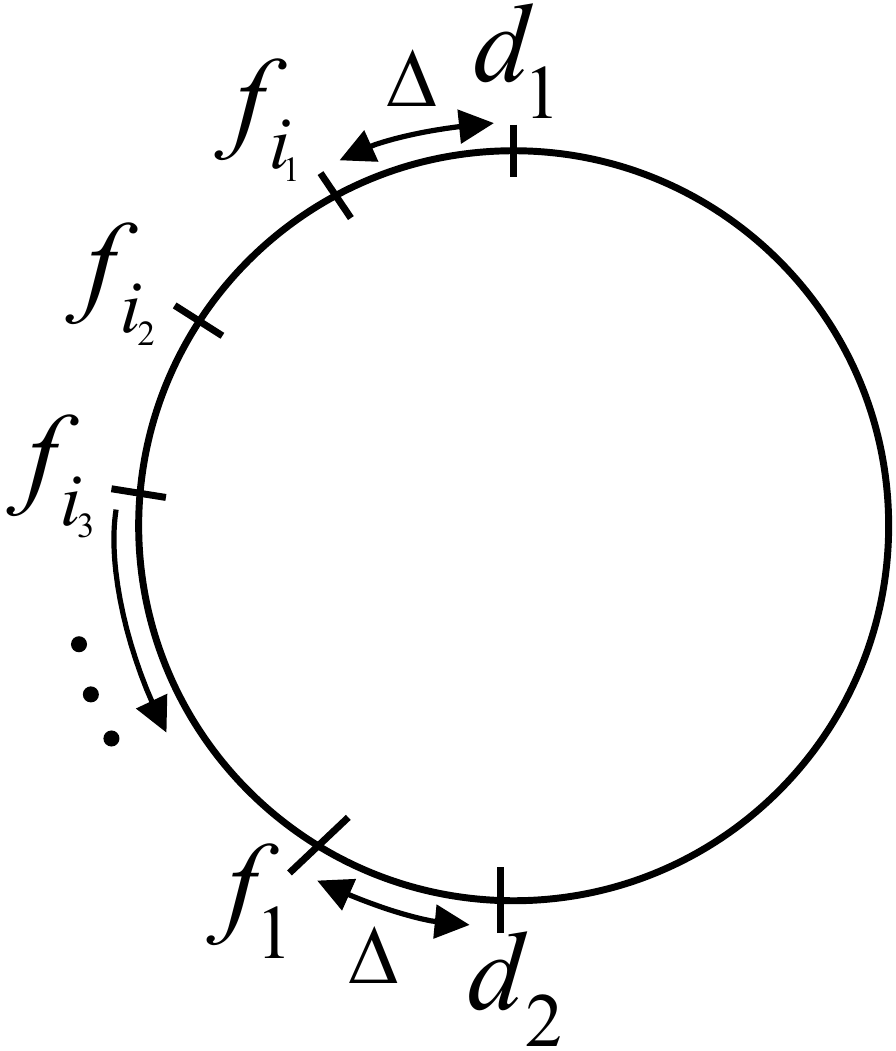}
\label{fig_suf_wwo}}
\hfil
\subfloat[With wrap around]{\includegraphics[width=1.44in]{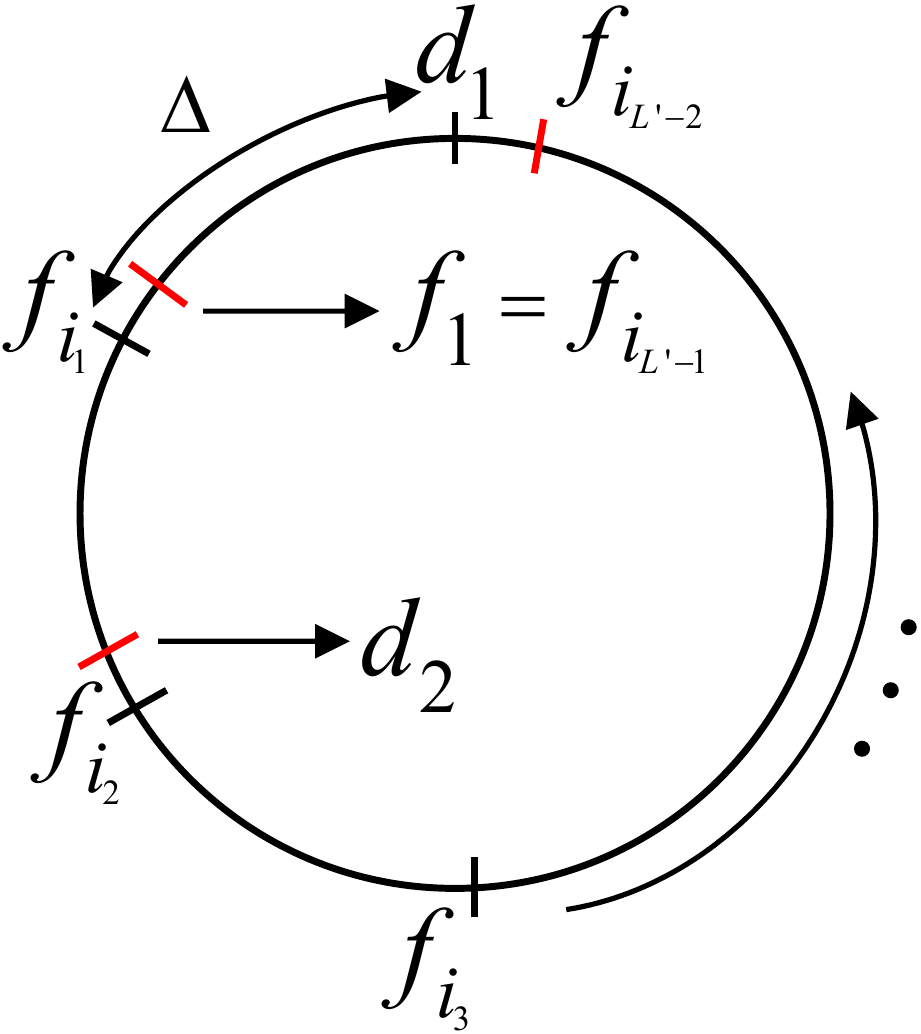}
\label{fig_suf_wwr}}}
\caption{One possible arrangement of elements of row 1 for Theorem~\ref{thm_main_girth8}}
\label{fig_str1}
\end{figure}

\subsection{Case 1}
\label{sec_subsec_case1_girth8}
Fig.~\ref{fig_str1} shows one of the two possible cases for the structure of the elements in $A_1 \cup \{d_2\}$ with respect to the circular representation.  In this case, the $L'+1$ elements $d_1, \dots, f_1, d_2$ form a single chain from $d_1$ to $d_2$, through $L'$ intervals of $\Delta=x_{L'+2}-x_{L'+1}$ points each, as shown in Fig.~\ref{fig_str1}. WLOG the chain begins at $d_1$ and progresses anticlockwise or else the two rows can be exchanged to yield this. Fig.~\ref{fig_str1}\subref{fig_suf_wwr} considers any possible wrap around.

\begin{lemma}
For a valid $G_8$ table that falls in the case illustrated by Fig.~\ref{fig_str1}, there exists a third row whose $L'$ elements are all different from the $L'+1$ elements in the chain $(d_1, f_{i_1}, f_{i_2}, \dots, f_{i_{L'-2}}, f_1, d_2)$, where $\{i_1, i_2, \dots, i_{L'-2}\} = \{2, 3, \dots, L'-1\}$.
\end{lemma}

\begin{IEEEproof}
Any row beyond the first two rows of the $G_8$ table relates to $A_1$ by an offset $\Delta' \ne \Delta$.   If $d_1$ or $d_2$ repeat in such a row then the $G_8$ table is not valid. With $ \{i_1, i_2, \dots, i_{L'-1}\}= \{1, 2, \dots, L'-1\}$ and with $i_{L'-1}=1$, we need to show that for a valid $G_8$ table that falls in Case 1, for any $k \ne \ell \in \{1, 2, \dots, L'-1\}, f_{i_k} + \Delta' \ne f_{i_\ell}$  and $d_1 + \Delta' \ne f_{i_\ell}$. 

Assume for a contradiction that $\exists ~k \ne \ell : f_{i_k} + \Delta' = f_{i_\ell}$. Define $n\Delta = \underbrace{\Delta+\Delta+\dots+\Delta}_{n \text{ times}}$, where $n$ is any non-negative integer.  If $n$ is negative, define $n\Delta=\underbrace{-\Delta-\Delta- \dots - \Delta}_{-n \text{ times}}$.  If $\ell > k$, then $f_{i_\ell} = f_{i_k} + (\ell - k)\Delta$ and hence $\Delta' = (\ell - k)\Delta$. Since $1 \le \ell - k < L'-1$ we can also obtain that $d_2=f_{i_{L'-{(\ell - k)}}}+(\ell - k)\Delta$. This shows that $d_2$ would be an element of the new row, yielding a contradiction. The same argument in the opposite direction will show that $d_1$ will repeat as $d_1 = f_{i_{k - \ell}}+\Delta'$ if $\ell < k$. Similarly, one can show that, if $d_1 + \Delta' = f_{i_\ell}$ then $d_2$ will repeat. 
\end{IEEEproof}

To summarize, there exist at least $(L'+1)+L'+(L'-3)=3L'-2$ distinct non-zero elements in a $G_8$ table that is valid and falls in Case 1, which means $N \ge 3L'-1$: $L'+1$ elements from $A_1 \cup \{d_2\}$, $L'$ elements in a third row and the term $L'-3$ appears from counting at least one distinct non-zero entry (on the diagonal) from each of the remaining rows.

\subsection{Case 2}
\label{sec_subsec_case2_girth8}
The situation where a single chain is not present within the set $A_1 \cup \{d_2\}$ is considered now. This is because, introducing only one new element when creating the second row from the first row, i.e.\ $d_2$, can also arise from the situation shown by the example in Fig.~\ref{fig_str2} (refer to the following description). 

\begin{figure}[h]
\centerline{{\includegraphics[width=2.6in]{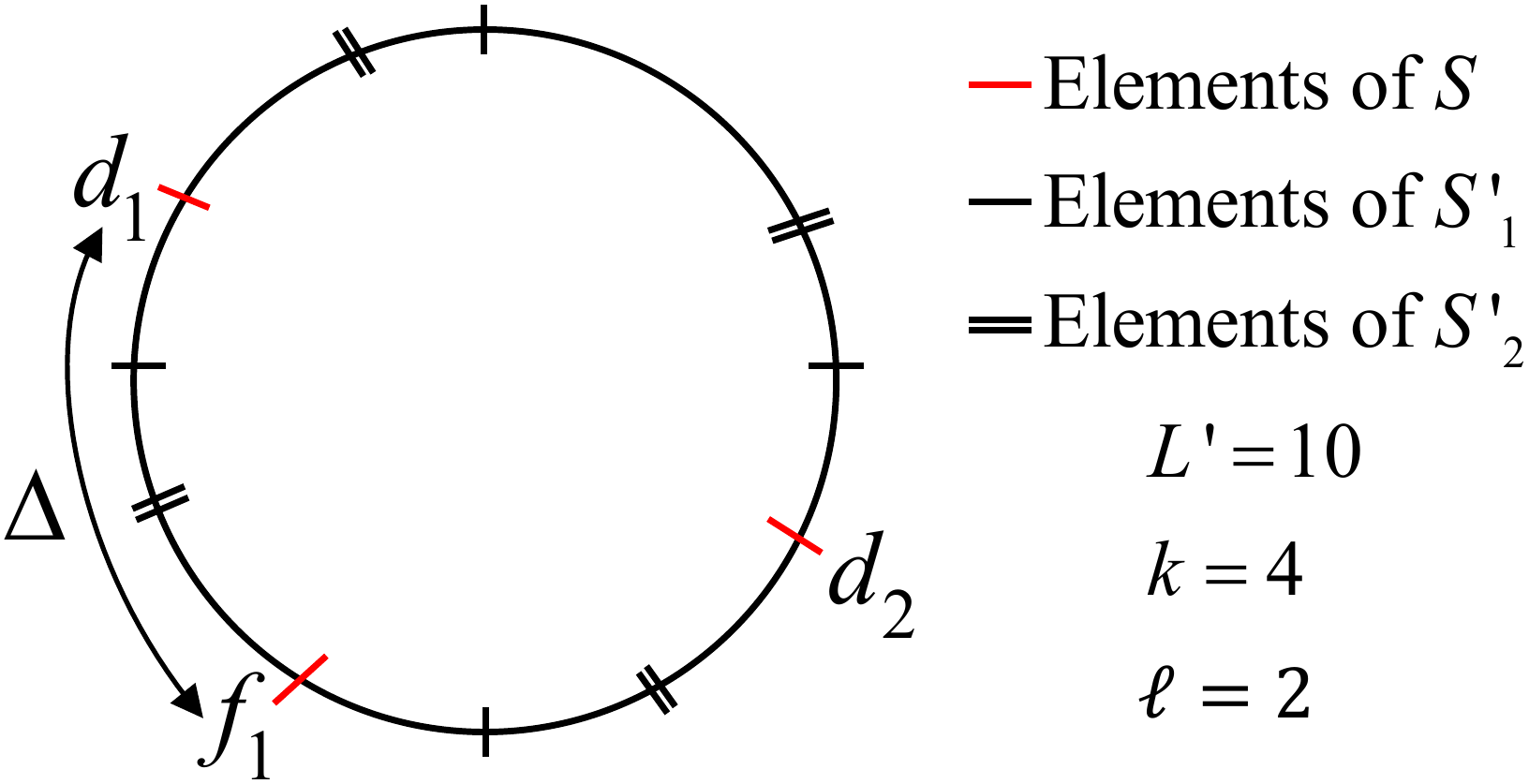}}}
\caption{Alternative arrangement of elements of row 1 for Theorem~\ref{thm_main_girth8}}
\label{fig_str2}
\end{figure}

The elements of $A_1 \cup \{d_2\}$ could be in $\ell+1$ disjoint sets:
\begin{enumerate}
\item Set $S$ with elements $\{d_1, \underbrace{f_{i_1}, \dots, f_{i_{L'-2-k\ell}}, f_1}_{L'-1-k\ell \text{ elements}},d_2\}$. 
\item $\ell$ other sets denoted $S'_j, 1 \le j \le \ell $ each comprising $k$ elements such that within each set adding $n\Delta, n \in \mathbb{Z}$ to any element yields another element within the set itself.
\end{enumerate}

While this case is introduced here, the proof that the theorem holds for it is given in the Appendix. This concludes the proof of Theorem~\ref{thm_main_girth8}. 
\end{IEEEproof}

Although Theorem~\ref{thm_main_girth8} only applies under specific constraints on the girth-8 table, we conjecture that the bound $N \ge 3L' - 1$ applies without these imposed constraints that $\exists~ i \ne j \in \{1, 2, \dots, L'\}$ such that $\left|A_i \cap A_j \right|$  equals 0 or $L'-1$.
\section{Conclusion}
\label{sec_conclusion}
This paper considers the problem of constructing $(3,L)$ quasi-cyclic low-density parity-check (LDPC) codes from complete protographs. An application of complete mappings from finite group theory provides explicit constructions of $(3,L)$ QC-LDPC codes that achieve girth $g=6$ with the minimum possible lifting factor of $L$ when $L$ is odd. Identifying the minimum lifting factor required to obtain a girth of $g \ge 8$ is posed as a problem in additive combinatorics using the construct of a girth-8 table ($G_8$ table).  An improved bound on the lifting factor  is obtained under certain constraints on the cardinality of girth-8-table row-set intersections.  We conjecture that this improved bound applies in general.
\bibliographystyle{IEEEtran}
\bibliography{IEEEabrv,mybib}
\appendix[Proof of Case 2]
\label{sec_appendix}
We prove here that Theorem~\ref{thm_main_girth8} holds for Case 2 which was introduced in Section~\ref{sec_subsec_case2_girth8}. The $\ell$ sets being referred to in Case 2 (from Fig.~\ref{fig_str2}) have the same number of elements, denoted $k$ here, or else adding $\Delta$ will create a new element for the second row, apart from $d_2$ which is already being created from $f_1 \in S$. To show that the theorem holds for this case, we focus on the $k \ell$ elements from the $\ell$ sets. For this case, a ``linear'' relationship within the elements of the $\ell + 1$ sets holds as follows. For the elements in $S$, 
\begin{align*}
d_1 + \Delta =& f_{i_1}, \\ 
f_{i_1} + \Delta =& f_{i_2}, \\
&\vdots \notag \\
f_{i_{L'-2-k\ell}} + \Delta =& f_1, \\
f_1 + \Delta =& d_2.
\end{align*}
For the elements in the $\ell$ sets $S'_j$, 
\begin{align}
\label{e11}
\forall x \in S'_j, 1 \le j \le \ell, x + n\Delta \in S'_j, \forall n \in \mathbb{Z}.
\end{align}
Observe that if the elements from the group $\mathbb{Z}/N$ are chosen for $A_1$ according to Case 2, then the following holds:
\begin{align}
\label{e9}
\forall x \in \mathbb{Z}/N, x+k\Delta = x.
\end{align}

Each column of the $G_8$ table has a diagonal element that appears only once in the table.  By adding offsets to the $k \ell$ elements in $\left(A_1 \cup \{d_2\}\right) \setminus S $ to obtain the $k \ell$ corresponding diagonal elements (and their respective rows), we have the following crucial observation. 

\begin{lemma}
For each column corresponding to an element in $\left(A_1 \cup \{d_2\}\right) \setminus S$, when obtaining a new diagonal element, at least $k$ new non-zero elements are obtained in the  corresponding row. Considering all $k \ell$ such rows, a total of $k^2 \ell$ distinct elements, that are different from the elements in $A_1 \cup \{d_2\}$, is guaranteed for any valid $G_8$ table that falls under Case 2. 
\end{lemma}

\begin{IEEEproof}
Consider $x_1 \in \left(A_1 \cup \{d_2\}\right) \setminus S$ and assume that $x_1 \in S'_{i_1}$ for some $1 \le i_1 \le \ell$. There is an offset $\Delta_1 \ne \Delta$ such that $x_1+\Delta_1 = d_{x_1}$, where $d_{x_1}$ is the diagonal element in the column containing $x_1$.  Note that the row containing $d_{x_1}$ also contains every element in $S'_{i_1}+\Delta_1=\{s+\Delta_1 : s \in S'_{i_1}\}$.  No element in $S'_{i_1}+\Delta_1$ appears in $A_1 \cup \{d_2\}$ as this would force either  $S'_{i_1}+\Delta_1 = S'_j, j\ne i_1$ so that $d_{x_1} \in A_1 \cup \{d_2\}$ or $S \subseteq S'_{i_1}+\Delta_1$ so that $d_1$ and $d_2$ appear in the row containing $d_{x_1}$ due to~\eqref{e11}.  Either of these results would lead to a $G_8$ table that is not valid.

Now consider a second element  $x_2 \in \left(A_1 \cup \{d_2\}\right) \setminus S$,  $x_2 \ne x_1$ and $x_2 \in S'_{i_2}$, where $1 \le i_2 \le \ell$ is not necessarily different from $i_1$. There is an offset $\Delta_2 \notin \{\Delta, \Delta_1\}$ such that $x_2+\Delta_2 = d_{x_2}$, where $d_{x_2}$  is the diagonal element in the column containing $x_2$. Note that the row containing $d_{x_2}$ also contains every element in $S'_{i_2}+\Delta_2$.  Following the same reasoning as with $x_1$, no element in $S'_{i_2}+\Delta_2$ appears in $A_1 \cup \{d_2\}$.  Also, $\left(S'_{i_1}+\Delta_1 \right) \cap \left( S'_{i_2}+\Delta_2\right)  = \emptyset$ or else $S'_{i_1}+\Delta_1 = S'_{i_2}+\Delta_2$ due to~\eqref{e11} and in particular $d_{x_2} \in S'_{i_1}+\Delta_1$ which would lead to a $G_8$ table that is not valid. 

Continuing by induction yields $k^2\ell$ distinct elements that are not in the first row and are different from $d_2$.
\end{IEEEproof}

Thus we have for any valid $G_8$ table in Case 2 that
\begin{align}
\label{e7}
N \ge L'+2+k^2 \ell,
\end{align}
where $L'+2$ arises from counting the elements in $A_1 \cup \{d_2, 0\}$. 

\begin{lemma}
In the context of Case 2, where $|S|=L'-k\ell + 1$,
\begin{align}
\label{e8}
L'-k\ell + 1 \le k.
\end{align}
\end{lemma}

\begin{IEEEproof}
Due to~(\ref{e9}).
\end{IEEEproof}

Case 2 is only possible when $L' \ge 5$, $k \ge 3$ and  $\ell \ge 1$. We consider two ranges for $k$ as follows:
If $k \ge \sqrt{2L'-3}$, then
\begin{align}
N \ge& L' + 2 + k^2 \ell \notag \\
   \ge& L' + 2 + k^2 \notag \\
   \ge& L' + 2 + 2L'-3 = 3L' - 1.
\end{align}
If $k < \sqrt{2L'-3}$, we first use~(\ref{e8}) in~(\ref{e7}) to get
\begin{align}
N \ge& L' + 2 + k^2 \ell \notag \\
    \ge& L' + 2 + k(L'-k + 1) \notag \\
    =& L' + 2 + kL' - k^2 + k,
\end{align}
which yields a quadratic expression in $k$ for every $L'$. This is concave in $k$ and it can be verified that the maximum of the right-hand side is obtained at $k_{\text{max}} = \frac{L'+1}{2}$. Under the supposition that $k < \sqrt{2L'-3}$, we can also trivially verify that $k < \sqrt{2L'-3} < k_{\text{max}}$ for $L' \ge 5$ and thus (by concavity) to minimize the right-hand side, we have to set $k$ to the smallest feasible value, which is $k = 3$. This yields
\begin{align}
N \ge& L' + 2 + 3L' - 9 + 3 \notag \\
   =& 4L'-4 >3L' - 1, \forall L' \ge 5,
\end{align}
which completes the proof for Case 2 and thus of Theorem~\ref{thm_main_girth8}.
\end{document}